\documentclass[twocolumn,aps,showpacsnofootinbib,epsfig]{revtex4}
\usepackage{amssymb}
\usepackage{amsmath}
\usepackage{graphics}
\usepackage{epsfig}
\usepackage[dvips]{color}
\usepackage{bm}
\usepackage{simplewick}

\newcommand{\be}{\begin{equation}}
\newcommand{\ee}{\end{equation}}
\newcommand{\bea}{\begin{eqnarray}}
\newcommand{\eea}{\end{eqnarray}}
\newcommand{\beas}{\begin{eqnarray*}}
\newcommand{\eeas}{\end{eqnarray*}}

\newcommand{\nn}{\nonumber}
\newcommand{\slsh}[1]{{\not \! #1}}

\begin{document}
\title{Local Gauge Transformation for the Quark Propagator in an SU(N) Gauge Theory}
\author{M. Jamil Aslam$^1$, A. Bashir$^2$ and L.X. Guti\'errez-Guerrero$^3$}
\affiliation{$^1$ Department of Physics, Quaid-i-Azam University,
Islamabad
45320, Pakistan. \\
$^2$ Instituto de F\1sica y Matem\'aticas, Universidad Michoacana
de San Nicol\'as de Hidalgo, Edificio C-3, Ciudad Universitaria,
Morelia, Michoac\'an 58040, M\'exico. \\
$^3$ Departamento de F\'isica, Universidad de Sonora, Boulevard
Luis Encinas J. y Rosales, Colonia Centro, Hermosillo, Sonora
83000, M\'exico. \\
}
\begin{abstract}

In an $SU(N)$ gauge field theory, the $n$-point Green functions,
namely, propagators and vertices, transform under the simultaneous
local gauge variations of the gluon vector potential and the quark
matter field in such a manner that the physical observables remain
invariant. In this article, we derive this intrinsically non
perturbative transformation law for the quark propagator within
the system of covariant gauges. We carry out its explicit
perturbative expansion till ${\cal O}(g_s^6)$ and, for some terms,
till ${\cal O}(g_s^8)$. We study the implications of this
transformation for the quark-anti-quark condensate, multiplicative
renormalizability of the massless quark propagator, as well as its
relation with the quark-gluon vertex at the one-loop order.
Setting the color factors $C_F=1$ and $C_A=0$,
Landau-Khalatnikov-Fradkin transformation for the abelian case of
quantum electrodynamics is trivially recovered.

\end{abstract}


 \maketitle

\section{Introduction}\label{I}

In gauge field theories, Green functions transform in a specific
manner under a variation of gauge. In quantum electrodynamics
(QED), these transformations carry the name of
Landau-Khalatnikov-Fradkin transformations (LKFT) due to their
first derivation by these authors in the canonical formulation of
QED,~\cite{LKFT:1956}. Later, these were also derived by Johnson
and Zumino through functional methods,~\cite{Zumino:1959}. LKFT
are non perturbative in nature just like
Ward-Fradkin-Green-Takahashi identities (WFGTI),~\cite{WGTI:1950}.
The former govern the behavior of a single Green function when
variation of covariant gauge parameter is performed, whereas, the
latter relate different Green functions under local gauge
transformations.

WFGTI follow from the so called Becchi-Rouet-Stora-Tyutin~(BRST)
symmetry. One can further enlarge these transformations by
modifying the Lagrangian without affecting the dynamics of the
theory. Now also transforming the gauge
parameter~\cite{Nielsen:1975}, one can arrive at extended Ward
identities, known as Nielsen identities~(NI). An advantage of the
NI over the conventional Ward identities is that $\partial/
\partial \xi$ becomes an integral part of the new relations involving Green
functions. This fact was exploited in~\cite{Lavelle:1995} to prove
the gauge independence of the fermion pole mass related to the
two-point Green functions.

Although LKFT and WFGTI are independent relations, one can show
that if a given WFGTI holds in one gauge, then if the Green
functions involved are transformed to another gauge under LKFT,
the WFGTI in that other gauge will be satisfied automatically and
need not be rechecked. In general, the rules governing LKFT are
far from simple. The fact that they are written in coordinate
space adds to their complexity. As a result, these transformations
have played a less significant and practical role in the study of
gauge theories as compared to WFGTI.

In perturbation theory, both the WFGTI and the LKFT are satisfied
at every order of approximation. This perturbative approach is
sufficient to study QED even at the highest energies so far
achieved in experiments. The only gauge theory whose non
perturbative understanding is relevant to fundamental interactions
within the standard model is quantum chromodynamics (QCD). In the
infrared domain of QCD, emergent phenomena like dynamical chiral
symmetry breaking (DCSB) and confinement are observed which cannot
be accessed through the usual perturbative tools. Their gauge
invariance needs to be established in any physically acceptable
formulation of QCD. The gauge identities which relate Green
functions in QCD are called Slavnov-Tayolr identities
(STI),~\cite{Slavnov-Taylor:1971}. These identities play a key
role in the proof of gauge-invariant renormalizability of QCD.

    However, generalized LKFT for QCD have not been derived.
Making sure that gauge invariance remains intact in the non
perturbative studies of QCD is a highly non-trivial matter.
Therefore, it is important to have all relevant tools at hand,
including the non-abelian version of LKFT. Algebraically, QCD is
not too different from any other $SU(N)$ theory for our purpose.
So we set out to derive this transformation for the quark
propagator in an arbitrary $SU(N)$ gauge theory. Later, we can
specialize to QCD by setting the color factors $C_F=4/3$, $C_A=3$
or even to QED for the choice of $C_F=1$ and $C_A$=0.

We have organized this article as follows. In Sect. II, we derive
the local gauge transformation relation for the quark propagator
in the coordinate space. We provide its explicit perturbative
expansion till ${\cal O}(g_s^6)$ and, in case of some terms, till
${\cal O}(g_s^8)$. Sect. III is devoted to the implications of
these relations in perturbation theory in momentum space,
discussing the relation of the generalized LKFT with
multiplicative renormalizability of the massless quark propagator.
Sect. IV details the role of the quark-gluon vertex in ensuring
the multiplicative renormalizability of the quark propagator and
hence its connection with the LKFT. In Sect. V, we present
conclusions and some possible avenues of future work.

\section{Quark Propagator Under Local ${\bm{SU(N)}}$
Gauge Transformations}

 Local $SU(N)$ gauge transformation for the quark field reads as~:
 \bea
 \psi _{i}(x) & \hspace{-0.2cm} \rightarrow  \hspace{-0.15cm} & \psi _{i}^{\prime }(x)
 \hspace{-0.1cm} = \hspace{-0.08cm}
 {\rm e}^{i [g_{s} \varphi _{a}(x)T_{a}]} \; \psi _{i}(x) \,,  \label{GTquarks}
 \eea
 where $\varphi_a(x)$ is an arbitrary free scalar field operator carrying
 color quantum number $a$ ($a=1,2, \cdots, N^2-1$), $g_s$ is the strong coupling constant and
 $T_a= \lambda_a/2$, $\lambda_a$ being the well-known Gell-Mann matrices for $N=3$.
 Following the work of Landau and Khalatnikov~\cite{LKFT:1956} for QED, we shall
 investigate how the quark Green function (quark propagator in the coordinate
 space) would transform under this gauge transformation. The quark
 Green function is defined as~:
 \bea
 i S^F_{ij} (x,x') \equiv i S^F_{ij} = \langle {\rm T} \{ \psi_{i}(x) \bar{\psi} _{j}(x') \} \rangle \;,
 \label{GFquarks}
 \eea
 where the angle brackets stand for the vacuum expectation value.
 Under  local $SU(N)$ local gauge transformation of
 Eq.~(\ref{GTquarks}), we have
 \bea
  i S^F_{ij}(x,x') \hspace{-0.1cm} \rightarrow \hspace{-0.13cm}
 \langle {\rm T} \{ {\rm e}^{i[g_{s} \varphi_{a}(x) T_{a}]}
  \psi_{i}(x) \hspace{-0.02cm} \bar{\psi} _{j}(x')
  {\rm e}^{-i[g_{s}\varphi _{b}(x') T_{b}]}
 \rangle.
 \eea
 Note that the color matrices act on the corresponding fermion
 fields. If we explicitly show the color indices, we can write~:
 \bea
   && \hspace{-0.5mm} (T_a \psi_i) \, (\bar{\psi}_j T_b) \; [ \; \Rightarrow \; ((T_a \psi_i)  \, (\bar{\psi}_j T_b))_{\sigma \mu}
   ] \nn \\
   \hspace{-2mm} && \hspace{-4mm} = (T_a)_{\sigma \lambda}  (\psi_i)_{\lambda} \;
   (\bar{\psi}_j)_{\nu} (T_b)_{\nu \mu}
   =  \left[ (T_a)_{\sigma \lambda} (T_b)_{\nu \mu} \right] \, (\psi_i)_{\lambda}
   (\bar{\psi}_j)_{\nu}. \nn
 \eea
 The vacuum expectation value yields a Kronecker delta-function
 for the color indices of the fermion fields, i.e., $\delta_{\lambda
 \nu}$. Thus we are left with (for the color part)
 \bea
  \left[ (T_a)_{\sigma \lambda} (T_b)_{\nu \mu} \right] \delta_{\lambda
  \nu} =  (T_a)_{\sigma \nu} (T_b)_{\nu \mu} = [ T_a \, T_b ]_{\sigma \mu}
  \nn\;.
 \eea
 The same will be true for any number of generators multiplying together.
 So we omit the explicit color  indices on the generating matrices as well
 as on the quark fields and
 indicate only the Lorentz indices of the quark fields, namely $i,j$.
 Since there are two fermion and two scalar fields, the only
 surviving contractions get factorized as~:
 \vspace{-4mm}
 \bea
   i S^F_{ij}(x,x') \hspace{-0.05cm} \rightarrow \hspace{-0.05cm}
   i S^{0F}_{ij}(x,x') \langle {\rm T} \{ {\rm e}^{i[g_{s} \varphi_{a}(x) T_{a}]}
  \, {\rm e}^{-i[g_{s}\varphi _{b}(x') T_{b}]}
 \rangle, \label{gauge-part} \vspace{3mm}
 \eea
where $S^{0F}_{ij}(x,x')$ stands for the Green function for quarks
for the particular case when the longitudinal part of the gluon's
Green function is zero (Landau gauge).
Note that this expression for a non-abelian $SU(N)$ theory is
inherently different from the corresponding abelian expression
which has no non-commuting operators in the exponentials and hence
can trivially be combined. In QCD, the non-commuting part of the
algebra (gluon self-interactions) introduces the color factor
$C_A$ which enters only at ${\cal O}(g_s^4)$ in the quark
propagator expansion. In the abelian approximation of QCD, i.e.,
$C_A=0$, the final result resembles that of QED with $C_F$ sitting
as an overall multiplication factor:
 \bea
 i S^F_{ij}(x,x')  = i S^{0F}_{ij}(x,x') \; {\rm e}^{ \, {g_s^2 } \,
 C_F \, \left[ i \Delta_F(x-x') - i \Delta_F(0) \right]}.
 \label{abelianLKF}
 \eea
On including the gluon self interactions, one can either use the
Bakker-Cambell-Hausdorff formula for combining the non-commuting
group generators or expand out the exponential, use the color
algebra to simplify the results and look for patterns in the
emerging series expansion to identify them with well-known
exponential expressions.

 The odd power terms in the scalar field are annihilated on taking
 the vacuum expectation value and hence always drop out. We employ the
 following notation for the Green functions associated with the
 scalar fields~:
 \bea
 i \Delta_{F} (x-x') \delta_{ab} &=& \langle 0 | {\rm T} \{ \varphi_{a}(x) \varphi _{b}(x') \} | 0 \rangle \;.
 \label{GFscalars}
 \eea
We confirm the abelian-like result, Eq.~(\ref{abelianLKF}) by
expanding out Eq.~(\ref{gauge-part}) till ${\cal O}(g_s^8)$. It is
done by employing the Wick theorem on the time ordered products
and evaluating all the contractions involved in the series. The
symmetry pattern emerging at every increasing order of
perturbation theory is:
 \bea
   C_F^n  &\Rightarrow& \frac{1}{n!} \left[ i \Delta_F(x-x') - i \Delta_F(0)
   \right]^n \,.
 \eea
 This pattern requires a careful identification of the $C_F$
 factors in general because both $C_F$ and $C_A$ involve the
 number of colors $N$. Additionally, the factor
 $[ i \Delta_F(x-x') - i \Delta_F(0)]$
 is also
 crucial to guarantee the gauge invariance of the chiral quark condensate
 as we shall see shortly.

 The quark gluon interaction as well as the gluon 3- and 4-point
 interactions introduce factors of $C_F- C_A/2$ (e.g., abelian type one-loop quark-gluon
 vertex) and $C_A$ (e.g., non-abelian one-loop quark-gluon
 vertex, as well as gluon and ghost loops)
 in an intricate manner at increasing order of perturbation
 theory, resulting in non trivial coefficients of $i
 \Delta_F(x-x')$, $i \Delta_F(0)$ and their overlapping terms.
 We have calculated these additional terms to ${\cal O}(g_s^8)$.
 We present some defining steps of the calculation till ${\cal
 O}(g_s^4)$. As the details can be cumbersome, we shall only
 give the final expression till ${\cal O}(g_s^6)$. We do not give
 all the explicit expressions at ${\cal O}(g_s^8)$ because these terms do not
 add new features to the general form of the LKFT unless we
 mention it otherwise.

 Let us calculate the term involving the exponentials~:

 \begin{widetext}
 \bea
 && \langle 0|  {\rm T}\{  {\rm e}^{i[g_{s} \varphi_{a}(x) T_{a}]} \, {\rm e}^{-i[g_{s}\varphi _{b}(x') T_{b}]} \} | 0
 \rangle \nn \\
 &=& \langle 0|  {\rm T} \Bigg\{ \left[ 1 + i g_s \varphi_{a}(x) T_{a} +
 \frac{(ig_s)^2}{2!} (\varphi_{a}(x) T_{a})^2 +
 \frac{(ig_s)^3}{3!} (\varphi_{a}(x) T_{a})^3 +
 \frac{(ig_s)^4}{4!} (\varphi_{a}(x) T_{a})^4  + \cdots \right]
 \nn \\
 && \hspace{0.7cm}  \times
 \left[ 1 - i g_s \varphi_{b}(x') T_{b} +
 \frac{(-ig_s)^2}{2!} (\varphi_{b}(x') T_{b})^2 +
 \frac{(-ig_s)^3}{3!} (\varphi_{b}(x') T_{b})^3 +
 \frac{(-ig_s)^4}{4!} (\varphi_{b}(x') T_{b})^4
 + \cdots \right] \Bigg\} | 0 \rangle \,. \nn
 \eea
 We keep terms till ${\cal O}(g_s^4)$. Note that only terms of the type:
 $\varphi^0, \varphi^2, \varphi^4$ survive. The linear and cubic terms in the scalar field
 drop out when averaged in the vacuum state~:
 \bea
 && \langle 0|  {\rm T}\{  {\rm e}^{i[g_{s} \varphi_{a}(x) T_{a}]} \, {\rm e}^{-i[g_{s}\varphi _{b}(x') T_{b}]} \} | 0
 \rangle \nn \\
 &=& \langle 0|  {\rm T} \Bigg\{  1 + g_s^2 \; [\varphi_{a}(x) \varphi_{b}(x')] \; T_{a} T_{b}
 - \frac{g_s^2}{2} \; [\varphi_{a}(x) \varphi_{b}(x)] \; T_{a} T_{b}
 - \frac{g_s^2}{2} \; [\varphi_{a}(x') \varphi_{b}(x')] \; T_{a}
 T_{b} \nn \\
 &+&
 \frac{g_s^4}{4!} \; [\varphi_{a}(x) \varphi_{b}(x) \varphi_{c}(x) \varphi_{d}(x)] \; T_{a}T_{b}T_{c}T_{d}
 + \frac{g_s^4}{4!} \; [\varphi_{a}(x') \varphi_{b}(x') \varphi_{c}(x') \varphi_{d}(x')] \;
 T_{a}T_{b}T_{c}T_{d} \nn \\
 &-& \frac{g_s^4}{3!} \; [\varphi_{a}(x) \varphi_{b}(x) \varphi_{c}(x) \varphi_{d}(x')] \; T_{a}T_{b}T_{c}T_{d}
 - \frac{g_s^4}{3!} \; [\varphi_{a}(x) \varphi_{b}(x') \varphi_{c}(x') \varphi_{d}(x')] \; T_{a}T_{b}T_{c}T_{d}
 \nn \\
 &+& \frac{g_s^4}{4} \; [\varphi_{a}(x) \varphi_{b}(x) \varphi_{c}(x') \varphi_{d}(x')] \; T_{a}T_{b}T_{c}T_{d} \;  +
 {\cal O}(g_s^6) \Bigg\} | 0 \rangle  \label{intermediate} \;.
 \eea
 Note that at ${\cal O}(g_s^4)$, quark
 propagator involves self interactions of gluons which bring
 in factors of $C_A$ (the quark gluon vertex will contribute factors of $C_F - C_A/2$ for the abelian diagram
 and $C_A$ for the non-abelian diagram including 3-point self gluon
 interactions). The ghost and gluon loop will also contribute a
 factor of $C_A$.
 These factors do not have a counterpart in QED as
 photons are not self-interacting. Therefore, we would like to detail the ${\cal O}(g_s^4)$
 calculation. We proceed to do it as follows~:
 \bea
 &&  \langle 0 | {\rm T} \{ \varphi_{a}(x) \varphi_{b}(x) \varphi_{c}(x) \varphi_{d}(x) \} | 0 \rangle
  \; T_a T_b T_c T_d  \nn \\
 &=&
  \Big[ \langle 0 |
 \contraction{}{\varphi}{_a(x)}{\varphi}
 \contraction{\varphi_a(x) \varphi_b(x)}{\varphi}{_c(x)}{\varphi}
 \varphi_a(x) \varphi_b(x) \varphi_c(x) \varphi_d(x)
 | 0 \rangle +  \langle 0 |
 \contraction{}{\varphi}{_a(x) \varphi_b(x)}{\varphi}
 \contraction[1.5ex]{\varphi_a(x)}{\varphi}{_b(x)\varphi_c(x)}{\varphi}
 \varphi_a(x) \varphi_b(x) \varphi_c(x) \varphi_d(x)
 | 0 \rangle +  \langle 0 |
 \contraction{}{\varphi}{_a(x) \varphi_b(x) \varphi_c(x)}{\varphi
 }
 \contraction[1.5ex]{\varphi_a(x)}{\varphi}{_b(x)}{\varphi}
 \varphi_a(x) \varphi_b(x) \varphi_c(x) \varphi_d(x)
 | 0 \rangle \Big]  \; T_a T_b T_c T_d  \nn \\
 &=& i \Delta_F(0) \; i \Delta_F(0) \left[ \delta_{ab} \delta_{cd} +
 \delta_{ac} \delta_{bd} + \delta_{ad} \delta_{bc} \right] \; T_a T_b T_c T_d
 = (i \Delta_F(0))^2 \left[ T_a T_a T_c T_c + T_a T_b T_a T_b + T_a T_b T_b T_a \right] \,. \nn
 \eea
 It is the middle term in the square bracket of the last term which
 brings in the non-abelian nature of QCD as two identical $T_i$ are not next
 to each other. Using the color identities collected in the
 appendix, we arrive at the following results:
 \bea
   \langle 0 | {\rm T} \{ \varphi_{a}(x) \varphi_{b}(x) \varphi_{c}(x) \varphi_{d}(x) \} | 0 \rangle
  \; T_a T_b T_c T_d  &=& (i \Delta_F(0))^2 \left[ 3 C_F^2 - \frac{1}{2} \, C_A \, C_F
  \right] \nn \\
   \langle 0 | {\rm T} \{ \varphi_{a}(x') \varphi_{b}(x') \varphi_{c}(x') \varphi_{d}(x') \} | 0 \rangle
  \; T_a T_b T_c T_d  &=& (i \Delta_F(0))^2 \left[ 3 C_F^2 - \frac{1}{2} \, C_A \, C_F
  \right] \nn \\
  \langle 0 | {\rm T} \{ \varphi_{a}(x) \varphi_{b}(x) \varphi_{c}(x) \varphi_{d}(x') \} | 0 \rangle
  \; T_a T_b T_c T_d  &=& (i \Delta_F(0)) \, (i \Delta_F(x-x')) \, \left[ 3 C_F^2 - \frac{1}{2} \, C_A \, C_F
  \right] \nn \\
    \langle 0 | {\rm T} \{ \varphi_{a}(x) \varphi_{b}(x') \varphi_{c}(x') \varphi_{d}(x') \} | 0 \rangle
  \; T_a T_b T_c T_d  &=& (i \Delta_F(0)) \, (i \Delta_F(x-x')) \, \left[ 3 C_F^2 - \frac{1}{2} \, C_A \, C_F
  \right] \nn \\
   \langle 0 | {\rm T} \{ \varphi_{a}(x) \varphi_{b}(x) \varphi_{c}(x') \varphi_{d}(x') \} | 0 \rangle
  \; T_a T_b T_c T_d  &=&  (i \Delta_F(0))^2    \,  C_F^2 + (i \Delta_F(x-x'))^2
   \left[ 2 C_F^2 - \frac{1}{2} \, C_A \, C_F
  \right] \,.
 \eea
 We can now simplify the ${\cal O}(g_s^4)$ terms in
 expression~(\ref{intermediate}) and collect the coefficients of
 $(i \Delta_F(0))^2$, $(i \Delta_F(x-x'))^2$ and $(i \Delta_F(0)) \,
 (i \Delta_F(x-x'))$ to arrive at the final expression to
 ${\cal O}(g_s^4)$~:
  \bea
  i S^F_{ij}(x,x') &=& i S^{0F}_{ij}(x,x') \Big[ {\rm e}^{ \, {g_s^2 } \, C_F \, \left[ i \Delta_F(x-x') - i \Delta_F(0)
 \right]}   \nn \\
   &-&
 \frac{g_s^4 \, C_A \, C_F}{24} \left\{ \left( i \Delta_F(x-x') - i \Delta_F(0) \right)
   \left( 3 i \Delta_F(x-x') - i \Delta_F(0) \right) \right\}
   + {\cal O}(g_s^6) \Big] \label{Ogs4-summed} \;.
 \eea
 As expected, we have an additional term proportional to $C_A$ at ${\cal
 O}(g_s^4)$. This term corresponds to the next to leading log term in the two-loop perturbative expansion
 of the massless quark propagator. Note that it has the desired factor $[i \Delta_F(x-x') - i
 \Delta_F(0)]$ which ensures chiral quark condensate is gauge
 invariant in QCD. We will not give the details of ${\cal
 O}(g_s^6)$ calculation. However, in the appendix, one can find
 the relevant identities, formulas and contractions to arrive at
 the following final result.
  \bea
  i S^F_{ij}(x,x') &=& i S^{0F}_{ij}(x,x') \Big[ {\rm e}^  {{g_s^2 } \, C_F \, \left[ i \Delta_F(x-x') - i \Delta_F(0)
 \right] }\nn \\
 && \hspace{-2cm} - \frac{g_s^4 \, C_A \, C_F}{(2!)(3! 2! 1!)} \left\{ \left[ i \Delta_F(x-x') - i \Delta_F(0)
 \right]
    \left[ 3 i \Delta_F(x-x') - i \Delta_F(0) \right] \right\} \left[ 1 + g_s^2 C_F \left( i \Delta_F(x-x') - i \Delta_F(0) \right) \right]  \nn \\
 && \hspace{-2cm}  + \frac{g_s^6 \, C_FC_A^2}{(1!) (4!3!2!1!)} \, \left[ i \Delta_F(x-x') - i \Delta_F(0) \right] \left[ 8(i \Delta_F(x-x'))^2-7(i \Delta_F(x-x'))(i \Delta_F(0))+(i
\Delta_F(0))^2 \right] + {\cal O}(g_s^8) \Big] .
 \eea
 This expression suggests a possible start of a new next to leading log series at ${\cal O}(g_s^4)$
 which might be summed up as follows:
   \bea
  i S^F_{ij}(x,x') &=& i S^{0F}_{ij}(x,x') \Big[ {\rm e}^  {{g_s^2 } \, C_F \, \left[ i \Delta_F(x-x') - i \Delta_F(0)
 \right] } \nn \\
 && \hspace{-2cm} - \frac{g_s^4 \, C_A \, C_F}{(2!)(3! 2! 1!)} \left\{ \left[ i \Delta_F(x-x') - i \Delta_F(0)
 \right]
    \left[ 3 i \Delta_F(x-x') - i \Delta_F(0) \right] \right\}  {\rm e}^  {{g_s^2 } \, C_F \, \left[ i \Delta_F(x-x') - i \Delta_F(0)
 \right] }  \nn \\
 && \hspace{-2cm}  + \frac{g_s^6 \, C_FC_A^2}{(1!) (4!3!2!1!)} \, \left[ i \Delta_F(x-x') - i \Delta_F(0) \right] \left[ 8(i \Delta_F(x-x'))^2-7(i \Delta_F(x-x'))(i \Delta_F(0))+(i
\Delta_F(0))^2 \right] + {\cal O}(g_s^8) \Big]
\label{GT-q-propagator}.
 \eea
 \end{widetext}
 We have verified the existence of this new exponential series till ${\cal O}(g_s^8)$.
 Eq.~(\ref{gauge-part}) is the $SU(N)$ modification of the
 LKF transformation for the fermion propagator in QED.
 Eq.~(\ref{GT-q-propagator}) is the ${\cal O}(g_s^6)$ expansion
 of this transformation (though the exponentials are infinite order).
 We summarize below some key remarks regarding this transformation:

 \begin{itemize}

 \item

 For QED, $C_F=1$, $g_s = e$, $C_A = 0$. On substituting these
 values, we recuperate the LKFT for the
 electron propagator~\cite{LKFT:1956}.

\item

 We find closed expression for the perturbative series
 in the color factor in the fundamental representation, i.e.,
 $C_F^n$. However, a closed expression for the perturbative series
 in the color factor in the adjoint representation, namely
 $C_A^n$, is a harder nut to crack as it has several sources.
 This work is in progress and will be reported elsewhere.

 \item

 Recall the definition of the chiral quark condensate and
 the final expression for the quark propagator in an arbitrary
 covariant gauge, Eq.~(\ref{GT-q-propagator}):
 \bea
    \langle \bar{\psi}  \psi \rangle_{\xi} &=& - {\rm Tr} \left[ S^F(x,x')
    \right]_{x'=x} \nn \\
  &=& - {\rm Tr} \left[ S^{0F}(x,x') \right]_{x'=x} = \langle \bar{\psi}  \psi \rangle_{0}
 \eea
 Hence {\em the chiral quark condensate is manifestly a gauge invariant
 quantity in any SU(N) theory}. It owes itself to the fact that
 at least one color factor $C_F$ appears at every order in perturbation theory
 in each term, and hence
 contributes an overall factor of $(i \Delta_F(x-x') - i
 \Delta_F(0))$ which guarantees the gauge invariance of the chiral
 quark condensate.

\item

 To the order we have carried out our calculation, we observe the
 factors:
 \bea
   C_A &\Rightarrow& \frac{1}{3! 2! 1!} \hspace{0.37cm} \left[ 3 i \Delta_F(x-x') - i \Delta_F(0)
   \right]\nn \\
   C_A^2 &\Rightarrow& \frac{1}{4! 3! 2! 1!} \; \Big [ 8\{i \Delta_F(x-x')\}^2 + \{ i \Delta_F(0) \}^2  \nn \\
   &&   \hspace{1.5cm} -7 \{i \Delta_F(x-x')\} \; \{ i \Delta_F(0) \}
   \Big] \,.
 \eea  \\
 We have verified that the first of this recurs at ${\cal
 O}(g_s^6)$ and ${\cal O}(g_s^8)$, and we conjecture it to become a coefficient of
 $ {\rm e}^ {{g_s^2 } \, C_F \, [ i \Delta_F(x-x') - i \Delta_F(0)
 ]}$.

 \end{itemize}

The local gauge transformation for the quark propagator has
verifiable consequence to all orders in perturbation theory. For
example, it implies multiplicative renormalizability of the
massless quark propagator which we set out to discuss in the next
section.

 \section{Gauge Transformation and Multiplicative Renormalizability}

 The connection of local gauge transformation for a charged fermion
 with its leading $(\alpha \xi)^n$ or sub-leading $\alpha^n \xi^{n-1}$
 perturbative expansion has been studied in detail for
 QED,~\cite{Kizilersu:2000}. For 4 space-time dimensions in massless
 QED, this expansion becomes a multiplicatively renormalizable power law for the
 wave function renormalization,~\cite{Raya:2002}. Something similar
 takes place in QCD but the gluon self-interactions introduce additional
 series as we now see.

 Two completely general Lorentz decompositions of quark propagator in momentum and
 coordinate space are, respectively:
 \bea
  S^F(p:\xi) &=& \frac{F(p;\xi)}{\slsh{p} - {\cal M}(p;\xi)} \,, \nn \\
  S^F(x:\xi) &=& \slsh{x} X(x;\xi) + Y(x;\xi)  \,,
 \eea
 where we have set $x'=0$ without the loss of generality and suppressed color indices.
 Let us start from the
 tree level quark propagator, i.e.,:
 \bea
   F(p;0) = 1 \;, \qquad {\cal M}(p;0) = 0 \,.  \label{Eq:part1}
 \eea
 We now proceed as follows. We Fourier transform this free quark
 propagator to the coordinate space and apply the gauge
 transformation law of Eq.~(\ref{GT-q-propagator}) to find
 the quark Green function $S^F(x;\xi)$ in an arbitrary covariant
 gauge. Its inverse Fourier transform yields the quark propagator
 back in momentum space. To check the consequences of the
 transformation law, we restrict ourselves to the one loop
 perturbation theory where gluon self interactions do not contribute
 to the quark propagator and we can set $C_A=0$.
 The Fourier transform of Eq.~(\ref{Eq:part1}) is:
 \bea
  X(x;0) = - \frac{1}{2 \pi^2 x^4} \,, \quad  Y(x;0)=0 \,.
 \eea
 From Eq.~(\ref{GFscalars}), we deduce:
 \bea
   i \Delta_F(x) = \xi \mu^{4-d} \int_0^{\infty} \frac{d^dp}{(2
   \pi)^d} \frac{{\rm e}^{-i p \cdot x}}{p^4} \, .
 \eea
 Here we have invoked $d$-dimensional integration to deal with
 ultraviolet divergences. Setting $d=4-2 \epsilon$,
 \bea
  i \Delta_F(x) = - \frac{\xi}{(4 \pi)^2} \left[ \frac{1}{\epsilon} + \gamma + 2 {\rm ln} (\mu x)
  + {\cal O}(\epsilon) \right] \,.
 \eea
 As $\Delta_F(0)$ is divergent, we introduce a cut-off scale $x_{\rm
 min}$ to write
 \bea
 i \Delta_F(x_{\rm min}) - i \Delta_F(x) = \lambda \, {\rm ln} \left( \frac{x^2}{x_{\rm min}}
 \right) \,,
 \eea
 where $\lambda = \xi/(4 \pi)^2$. Thus the LKF transformation
 yields:
 \bea
   S^F(x; \xi) &=& - \frac{\slsh{x}}{2 \pi^2 x^4} \,
   \left( \frac{x^2}{x_{\rm min}}  \right)^{\nu} \,,
 \eea
 where $\nu = C_F \, \alpha_s \xi/(4 \pi)$ and
 $\alpha_s = g_s^2/(4 \pi)$.
 Its inverse Fourier
 transform yields the quark propagator in momentum space
 \bea
  S^F(p;\xi) = \int d^4x {\rm e}^{i p \cdot x} \, S^F(x;\xi) \,.
 \eea
 Thus
 \bea
  F(p;\xi) = \frac{1}{2^{2 \nu}} \,
  \frac{\Gamma(1-\nu)}{\Gamma(2+\nu)} (p^2 x_{\rm min}^2)^{\nu} \,.
 \eea
 Note that in a cut-off regularization $x_{\rm min} \propto
 1/\Lambda$. Recall from the previous section that the requirement
 of multiplicative renormalizability implies
 \bea
  \frac{F_R(p^2/\mu^2;\xi)}{F_R(k^2/\mu^2;\xi)} =
  \frac{F(p^2/\Lambda^2;\xi)}{F(k^2/\Lambda^2;\xi)}\,,
 \eea
 where the notation $F(p^2/\Lambda^2;\xi) \equiv
 F(p;\xi)$ has been used for the sake of clarity. We can choose
 $k^2=\mu^2$ and impose the renormalization condition of
 restoring the tree level quark propagator at sufficiently
 large renormalization scale $\mu^2$. That is to say:
 \bea
     F_R(k^2/\mu^2;\xi)|_{k^2=\mu^2} = 1 \,.
 \eea
 Thus
 \bea
  F_R(p^2/\mu^2;\xi) = {\cal Z}_2^{-1}(\mu^2;\Lambda^2)
  F(p^2/\Lambda^2;\xi) \;,
 \eea
 where
 \bea
  {\cal Z}_2(\mu^2;\Lambda^2) = F(\mu^2/\Lambda^2;\xi) \,.
 \eea
 Hence the LKFT ensure multiplicative renormalizability of the
 quark propagator. Taking into account the gluon self interactions
 would imply a more involved exponent $\nu$ beyond the leading
 order. The general structure of the LKFT and
 multiplicative renormalizability suggests the following form of
 the exponent:
 \bea
  \nu = f_0 C_F \, \alpha_s  + f_1^a C_F^2 \alpha_s^2 + f_1^b C_F C_A \alpha_s^2
  + \cdots \,,
 \eea
 where $f_0,f_1^a,f_1^b \cdots$ are determined through
 perturbative series of the quark propagator or its LKFT.
 For example, $f_0 = \xi/ (4 \pi)$.
 These contributions, in addition to being connected to the one
 loop quark and gluon propagators, are also intricately related to the
 quark-gluon vertex beyond the tree level which dictates these
 corrections through the quark gap equation. This is what we opt to
 discuss in the next section because any non perturbative construction
 of the quark-gluon vertex must ensure multiplicative renormalizability
 of the quark propagator. It becomes all the more relevant for the quenched
 version of the theory.

\section{Quark-Gluon Vertex and LKF Transformations}

From the works in QED, we already know that the intricate
structure of the quark gluon vertex dictates multiplicative
renormalizability of the charged fermion and hence ensures LKF
transformations for the 2-point function are satisfied. Brown and
Dorey,~\cite{Brown:1991}, argue that an arbitrary construction of
the electron-photon vertex does not satisfy the requirement of the
multiplicative renormalizability. It was realized that neither the
bare vertex nor the Ball-Chiu-vertex,~\cite{Ball:1980}, which
satisfies the WFGTI, were good enough to fulfill the demands of
multiplicative renormalizability. Since then, starting from the
pioneering work by Curtis and Pennington,~\cite{Curtis:1993},
there have been improved attempts to incorporate the implications
of LKFT in constructing a reliable electron-photon vertex
ansatz,~\cite{Bashir:1994}. The need for the same in QCD was
realized in the work by Bloch, who constructs a model truncation
which preserves multiplicative renormalizability, and reproduces
the correct leading order perturbative behavior through assuming
non-trivial cancellations involving the full quark-gluon vertex in
the quark self-energy loop,~\cite{Bloch:2002}.

Note that the quark propagator beyond ${\cal O}(g_s^2)$ involves
gluon self interactions. These interactions introduce the color
factor $C_A$ in the adjoint representation. For example, one-loop
calculation of the transverse quark-gluon vertex, defined by
$(k-p)_{\mu} \Gamma^{\mu}(k,p)$=0, where $k$ and $p$ are the
incoming and outgoing fermion momenta, respectively, for the large
separation of quark momenta scales, i.e., $k^2 >> p^2$,
reveals,~\cite{xiomara:2015}:
 \bea
  \Gamma^{\mu}_{a}(k,p) &=& \left(C_F - \frac{1}{2} C_A \right) \,
  \frac{\alpha_s \xi}{8 \pi k^2} \; {\rm log} \left( \frac{p^2}{k^2} \right) \,
  T^{\mu}(k,p) \,, \nn \\
  \Gamma^{\mu}_{n}(k,p) &=& C_A \,
  \frac{\alpha_s (3 \xi -1)}{64 \pi k^2} \; {\rm log} \left( \frac{p^2}{k^2} \right) \,
  T^{\mu}(k,p) \,,
 \eea
 where the subscripts $a$ and $n$ correspond to the abelian and
 non-abelian diagram, respectively. Vector $T^{\mu}(k,p)$ is
 defined as $
   T^{\mu}(k,p) = k^2 \gamma^{\mu} - k^{\mu} \slsh{k}$. Note that
   one expect terms of the type $\alpha_s \xi^2$ from the
   non-abelian diagram but these terms identically cancel out in this limit. The
   gauge parameter dependent piece of the quark gluon vertex would
   result in the following type of ${\cal O}(g_s^4)$ terms in the
   wave-function renormalization~:
   \bea
   F(p^2/\mu^2;\xi) &=&
   1 + \alpha_s \, f_0 \, C_F \, {\rm ln} \left( \frac{p^2}{\mu^2} \right)  \nn \\
   && \hspace{-2.5cm} +
   \alpha_s^2 \Bigg[ \frac{f_0^2 C_F^2}{2} \, {\rm ln}^2 \hspace{-.0cm} \left( \frac{p^2}{\mu^2}
   \right)   + C_F \, ( f_1^a C_F + f_1^b C_A )  \,  {\rm ln} \left( \frac{p^2}{\mu^2} \right)  \Bigg] \nn  \,.
   \eea
   Note that to this
   order, the $C_A$ terms would also be contributed by the gluon
   and ghost loop correction to the gluon propagator in the quark
   gap equation.

\section{Conclusion}

We derive the generalized LKFT for the quark propagator in an
$SU(N)$ gauge field theory. We then expand this expression under
local variation of gauge~(\ref{gauge-part}) to ${\cal O}(g_s^6)$
and, for some terms, to ${\cal O}(g_s^8)$. We find a closed
expression for the perturbative series $C_F^n C_A^0$ and
$C_F^{n-1} C_A$. We expect the same for terms higher in the order
of $C_A$. However, as the color factor $C_A$ is present in all
different interactions including gluon 3- and 4-point self
interactions, interactions involving ghosts as well as the
quark-gluon interactions beyond one loop, it is not
straightforward to disentangle different series in closed forms.
However, work is in progress in this direction and we plan to
report it elsewhere. Due to the explicit presence of the overall
factor of $(i \Delta_F(x-x') - i \Delta_F(0))$, the gauge
invariance of the chiral quark condensate is guaranteed just as in
QED,~\cite{raya:2007}.

In the abelian approximation, we carry out the Fourier transform
and obtain a multiplicatively renormalizable power law for the
massless quark propagator . The general structure of the LKFT and
the quark-gluon interactions at one loop order allow us to foresee
the pattern of $C_A$ terms of the non-abelian origin, which appear
at the next to leading order level in the perturbative expression
for the massless quark propagator.

The generalized LKFT for QCD also provide us with information on
how gluon and ghost propagators as well as the three and four
point vertices transform under the local variation of
gauge. This work is in progress and will be reported elsewhere. \\

\noindent {\bf Acknowledgments:} This work has been supported by
CIC-UMICH and CONACyT grants. We thank Prof. R. Delbourgo and
Prof. M.R. Pennington for their helpful comments. We dedicate this
article to the memory of late Prof. Riazuddin with whom initial
ideas of the work were
discussed. \\
\vspace{-0.3cm}

\noindent {\bf Appendix:} \\

\vspace{-0.3cm}

\noindent The details and identities of color algebra can be found
in different textbooks. However, for the sake of completeness and
for a quick reference, we list most of the identities we used. The
starting commutation and anti-commutation relations are:
\begin{eqnarray*}
\left[ T_{b},T_{c}\right]  = if_{bcp}T_{p} \,, \quad  \left\{
T_{b},T_{c}\right\}  =\frac{1}{N}\delta
_{bc}+d_{bcp}T_{p} \,. \nn \\
\end{eqnarray*}
Similarly, we use the identities:
\begin{eqnarray*}
 && \hspace{-0.4cm} f_{aab} =0 \,, \; \; d_{aab} = 0 \,, \; \; f_{bcp}f_{bcq} = N\delta _{pq} \,, \; \; \delta _{pp} =N^{2}-1  \\
 && \hspace{-0.4cm}  d_{bcp}d_{bcq} =\left( N-\frac{4}{N}\right) \delta _{pq} \,, \quad \frac{1}{N} =-2\left( C_{F}-\frac{1}{2}C_{A}\right) \, .
\end{eqnarray*}%
Now some simple products of two or three $SU(N)$ generators can be
written as:
\begin{eqnarray*}
 && T_{a}T_{a} = C_{F} \,, \; \; T_{a}T_{b}T_{a} =\left( C_{F}-\frac{1}{2}C_{A}\right) T_{b} \, , \\
 && i f_{abc} \; T_c  T_a T_b = - \frac{1}{2} \, C_A \, C_F \;.
\end{eqnarray*}
Moreover,
 \bea
 T_a T_b T_a T_b &=&  C_F (C_F - \frac{1}{2} C_A) \, ,\nn \\
 T_a T_b T_c T_a T_b T_c &=& C_F (C_F - C_A) (C_F - \frac{1}{2}
 C_A)\, ,
 \nn \\
 T_a T_b T_c T_b T_a T_c &=&  C_F (C_F - \frac{1}{2} C_A)^2 \,, \nn \\
 T_a T_b T_a T_c T_b T_c &=& C_F (C_F - \frac{1}{2} C_A)^2  \,, \nn \\
 T_a T_b T_c T_a T_c T_b &=& C_F (C_F - \frac{1}{2} C_A)^2  \,, \nn \\
 T_a T_b T_c T_b T_c T_a &=& C_F^2 (C_F - \frac{1}{2} C_A)  \,.\nn
 \eea

 \begin{widetext}
 \bea
 (i \Delta_F(0)) \delta_{ab}[\langle 0 | {\rm T} \{ \varphi_{c}(x) \varphi_{d}(x) \varphi _{e} (x) \varphi_ {f} (x) \} | 0 \rangle] T_a T_b T_c T_d T_e T_f
 &=&
 (i \Delta_F(0))^3[C_F^2 \, ( 3C_F-\frac{1}{2}C_A )] ,\nn\\
 (i \Delta_F(0)) \delta_{ac}[\langle 0 | {\rm T} \{ \varphi_{b}(x) \varphi_{d}(x) \varphi _{e} (x) \varphi_ {f} (x) \} | 0 \rangle] T_a T_b T_c T_d T_e T_f
 &=&
 (i \Delta_F(0))^3[ C_F \, (C_F-\frac{1}{2}C_A)(3C_F-\frac{1}{2}C_A)] ,\nn\\
 (i \Delta_F(0)) \delta_{ad}[\langle 0 | {\rm T} \{ \varphi_{b}(x) \varphi_{c}(x) \varphi _{e} (x) \varphi_ {f} (x) \} | 0 \rangle] T_a T_b T_c T_d T_e T_f
 &=&
 (i \Delta_F(0))^3[C_F^3+C_F(C_F-\frac{1}{2}C_A)(2C_F-\frac{3}{2}C_A)] ,\nn\\
 (i \Delta_F(0)) \delta_{ae}[\langle 0 | {\rm T} \{ \varphi_{b}(x) \varphi_{c}(x) \varphi _{d} (x) \varphi_ {f} (x) \} | 0 \rangle] T_a T_b T_c T_d T_e T_f
 &=&
 (i \Delta_F(0))^3[ C_F \, (C_F-\frac{1}{2}C_A)(3C_F-\frac{1}{2}C_A)] ,\nn\\
(i \Delta_F(0)) \delta_{af}[\langle 0 | {\rm T} \{ \varphi_{b}(x)
\varphi_{c}(x) \varphi _{d} (x) \varphi_ {e} (x) \} | 0 \rangle]
T_a T_b T_c T_d T_e T_f
 &=&
(i \Delta_F(0))^3[C_F^2 \, (3 C_F-\frac{1}{2}C_A)] ,\nn\\
 (i \Delta_F(x-x')) \delta_{ab}[\langle 0 | {\rm T} \{ \varphi_{c}(x') \varphi_{d}(x') \varphi _{e} (x') \varphi_ {f} (x') \} | 0 \rangle] T_a T_b T_c T_d T_e T_f
 &=&
 (i \Delta_F(0))^2(i \Delta_F(x-x'))
   [C_F^2 \, ( 3C_F-\frac{1}{2}C_A )] ,\nn\\
 (i \Delta_F(x-x')) \delta_{ac}[\langle 0 | {\rm T} \{ \varphi_{b}(x') \varphi_{d}(x') \varphi _{e} (x') \varphi_ {f} (x') \} | 0 \rangle] T_a T_b T_c T_d T_e T_f
 &=&
 (i \Delta_F(0))^2(i \Delta_F(x-x')) \nn \\
 &\times& [ C_F \, (C_F-\frac{1}{2}C_A)(3C_F-\frac{1}{2}C_A)] ,\nn\\
 (i \Delta_F(x-x')) \delta_{ad}[\langle 0 | {\rm T} \{ \varphi_{b}(x') \varphi_{c}(x') \varphi _{e} (x') \varphi_ {f} (x') \} | 0 \rangle] T_a T_b T_c T_d T_e T_f
 &=&
 (i \Delta_F(0))^2(i \Delta_F(x-x')) \nn \\
 &\times& [C_F^3+C_F(C_F-\frac{1}{2}C_A)(2C_F-\frac{3}{2}C_A)] ,\nn\\
 (i \Delta_F(x-x')) \delta_{ae}[\langle 0 | {\rm T} \{ \varphi_{b}(x') \varphi_{c}(x') \varphi _{d} (x') \varphi_ {f} (x') \} | 0 \rangle] T_a T_b T_c T_d T_e T_f
 &=&
 (i \Delta_F(0))^2(i \Delta_F(x-x')) \nn \\
 &\times& [C_F (C_F-\frac{1}{2}C_A)(3C_F-\frac{1}{2}C_A)] ,\nn\\
(i \Delta_F(x-x')) \delta_{af}[\langle 0 | {\rm T} \{
\varphi_{b}(x') \varphi_{c}(x') \varphi _{d} (x') \varphi_ {e}
(x') \} | 0 \rangle] T_a T_b T_c T_d T_e T_f
 &=&
 (i \Delta_F(0))^2(i \Delta_F(x-x')) [C_F^2(3 C_F-\frac{1}{2}C_A)]
 \nn\\
 (i \Delta_F(0)) \delta_{ab}[\langle 0 | {\rm T} \{ \varphi_{c}(x') \varphi_{d}(x') \varphi _{e} (x') \varphi_ {f} (x') \} | 0 \rangle] T_a T_b T_c T_d T_e T_f
 &=&
 (i \Delta_F(0))^3  [C_F^2( 3C_F-\frac{1}{2}C_A) ] ,\nn\\
 (i \Delta_F(x-x')) \delta_{ac}[\langle 0 | {\rm T} \{ \varphi_{b}(x) \varphi_{d}(x') \varphi _{e} (x') \varphi_ {f} (x') \} | 0 \rangle] T_a T_b T_c T_d T_e T_f
 &=&
 (i \Delta_F(0))(i \Delta_F(x-x'))^2  \nn \\
 &\times& [ C_F (C_F-\frac{1}{2}C_A)(3C_F-\frac{1}{2}C_A)] ,\nn\\
 (i \Delta_F(x-x')) \delta_{ad}[\langle 0 | {\rm T} \{ \varphi_{b}(x) \varphi_{c}(x') \varphi _{e} (x') \varphi_ {f} (x') \} | 0 \rangle] T_a T_b T_c T_d T_e T_f
 &=&
 (i \Delta_F(0))(i \Delta_F(x-x'))^2 \nn \\
 &\times& [C_F^3+C_F(C_F-\frac{1}{2}C_A)(2C_F-\frac{3}{2}C_A)] ,\nn\\
 (i \Delta_F(x-x')) \delta_{ae}[\langle 0 | {\rm T} \{ \varphi_{b}(x) \varphi_{c}(x') \varphi _{d} (x') \varphi_ {f} (x') \} | 0 \rangle] T_a T_b T_c T_d T_e T_f
 &=&
 (i \Delta_F(0))(i \Delta_F(x-x'))^2 \nn \\
 &\times& [C_F (C_F-\frac{1}{2}C_A)(3C_F-\frac{1}{2}C_A)] ,\nn\\
(i \Delta_F(x-x')) \delta_{af}[\langle 0 | {\rm T} \{
\varphi_{b}(x) \varphi_{c}(x') \varphi _{d} (x') \varphi_ {e} (x')
\} | 0 \rangle] T_a T_b T_c T_d T_e T_f
 &=&
 (i \Delta_F(0))(i \Delta_F(x-x'))^2[C_F^2(3 C_F-\frac{1}{2}C_A)] \nn.
 \eea

\end{widetext}

\end{document}